\documentclass[]{aa}
\usepackage[utf8]{inputenc}
\usepackage[english]{babel}
\usepackage{graphicx}
\usepackage{graphbox}
\usepackage{xcolor}
\usepackage[export]{adjustbox}
\usepackage{soul}
\usepackage{booktabs} 
\usepackage{amssymb,amsmath}
\usepackage{array}
\usepackage{siunitx}
\usepackage{textgreek}
\usepackage{natbib}
\usepackage[varg]{txfonts}
\usepackage[normalem]{ulem}
\usepackage{url}
\usepackage[unicode=true]{hyperref}
\bibpunct{(}{)}{;}{a}{}{,} 
\hypersetup{
    colorlinks=true,
    linkcolor={blue!50!black},
    citecolor={blue!50!black}, 
    urlcolor={blue!80!black}   
}
\usepackage{etoolbox}

\begin{document}

\title{Modelling O-star astrospheres with different relative speeds between the ISM and the star: 2D and 3D MHD model comparison}
\titlerunning{O-star astrosphere models with different $u_{\mathrm{ISM}}$ in 2D and 3D}
\author{L. R. Baalmann\inst{\ref{tp4}} \and K. Scherer\inst{\ref{tp4},\ref{pci}} \and J. Kleimann\inst{\ref{tp4}} \and H. Fichtner\inst{\ref{tp4},\ref{pci},\ref{rapp}} \and D. J. Bomans\inst{\ref{pci},\ref{rapp},\ref{astro}} \and K. Weis\inst{\ref{astro}}}
\institute{Ruhr-Universität Bochum, Institut für Theoretische Physik IV, 44780 Bochum, Germany\label{tp4}\\
e-mail: \texttt{lb@tp4.rub.de} \and
Ruhr-Universität Bochum, Research Department, Plasmas with Complex Interactions, 44780 Bochum, Germany\label{pci} \and
Ruhr Astroparticle and Plasma Physics (RAPP) Center, 44780 Bochum, Germany\label{rapp} \and 
Ruhr-Universität Bochum, Astronomisches Institut, 44780 Bochum, Germany\label{astro}}
\date{Received 11 March 2022 / Accepted 21 April 2022}
\abstract{State of the art simulations of astrospheres are modelled using three-dimensional (3D) magnetohydrodynamics (MHD). An astrospheric interaction of a stellar wind (SW) with its surrounding interstellar medium (ISM) can only generate a bow shock if the speed of the interstellar inflow is higher than the fast magnetosonic speed.}{The differences of astrospheres at differing speeds of the ISM inflow are investigated, and the necessity of the third dimension in modelling is evaluated.}{The model astrosphere of the runaway O-star \textlambda~Cephei is computed in both two-\ and three-dimensional MHD at four different ISM inflow speeds, one of which is barely faster (superfast) and one of which is slower (subfast) than the fast magnetosonic speed.}{The two-dimensional (2D) and 3D models of astrospheres with ISM inflow speeds considerably higher than the fast magnetosonic speed are in good agreement. However, in 2D models, where no realistic SW magnetic field can be modelled, the downwind structures of the astrospheres vacillate. Models where hydrodynamic effects are not clearly dominant over the magnetic field show asymmetries, thus necessitating a 3D approach. The physical times of simulations of astrospheres with slow ISM inflows can swiftly exceed the lifetime of the corresponding star. A hitherto unobserved structure has been found downwind of the astrotail in the subfast 3D model.}{} 
\maketitle

\section{Introduction}\label{sec:intro}

Stellar bow shocks occur when the relative motion between a star and its environment is super(fast magneto)sonic. This speed criterion can be met either by a star at rest experiencing an ambient flow, which is referred to as the weather vane scenario \citep{povich+2008}, or, more commonly, by a runaway star moving through the interstellar medium \citep[ISM; e.g.][]{vanburen+1995}. While a wide variety of stellar objects can produce bow shocks \citep[e.g.][]{vanburen+1988}, for example, red supergiants and AGB stars \citep{cox+2012}, a significant number of bow shock observations are associated with OB stars \citep[e.g.][and references therein]{kobulnicky+2017}.

The interactions between a respective star's stellar wind (SW) and its surrounding ISM are typically modelled numerically, using two-dimensional (2D) hydrodynamics \citep[HD; e.g.][and references therein]{matsuda+1989,comeron+1998,scherer+2016}, 2D magnetohydrodynamics \citep[MHD;][]{vanmarle+2014,meyer+2017}, or, more recently, three-dimensional (3D) MHD \citep[e.g.][and references therein]{scherer+2020,meyer+2021,fraternale+2021}. From a computational perspective, the runaway and the weather vane scenarios are generally indistinguishable.

The interaction region of the SW and the ISM consists of three important features overall, all of which are MHD discontinuities \citep[cf., e.g.][Sects.~20.2~\&~20.3]{goedbloed+2010}: the bow shock (BS), at which the ISM inflow is decelerated from faster than a characteristic speed to slower than a characteristic speed; the astropause (AP), which separates the stellar and interstellar fluids; and the termination shock (TS) and Mach disk (MD), which are the counterparts of the BS for the SW. In MHD, the characteristic speed is the fast magnetosonic speed, $v_{\mathrm{f}}$, whereas in HD, it is the sonic speed, $v_{\mathrm{c}}$. If the ISM inflow is not faster than the characteristic speed, namely, if it is subfast or subsonic, no BS occurs. The BS, AP, TS, and MD are all surfaces of infinitesimal thickness; the BS, TS, and MD are genuine MHD shocks, whereas the AP is a tangential discontinuity. The BS is shaped like a paraboloid; its apex is pointed towards the direction from which the inflow comes, which is called the upwind direction, and lies in front of the star. The AP is bullet-shaped; its apex lies in front of the star but downwind from the BS, that is, in between the star and the BS. The TS is shaped like a stylised tulip (see, e.g.\ Figs.~\ref{fig:lcep_vism_2d}~\&~\ref{fig:lcep_vism_3d}) with a similar orientation and position of its apex, downwind from the AP. The MD is in good approximation shaped like a spherical cap and it serves as the counterpart to the TS in the downwind direction. Its apex is pointed downwind and it meets the TS at a roughly circular line that is, for historical reasons, known as the triple point \citep[TP;][]{bleakneytaub1949}. From the TP, another tangential discontinuity (TD), which is roughly cylindrical in shape \citep[e.g.][Sect.~4.1]{scherer+2016}, extends into the downwind direction. 

The region between the TS\,\&\,MD and the AP is referred to as the inner astrosheath, while the one between the AP and the BS is known as the outer astrosheath. Some authors, such as \citet{zank2015}, refer to these regions as the astrosheath and the very local interstellar medium (VLISM), respectively; the outer astrosheath is occasionally also referred to as the bow shell \citep[e.g.][]{henney+2019a}. What is often called a bow shock in observational studies is typically emission from the outer astrosheath \citep{baalmann+2020}. The region inside the TD is called the astrotail, though this term often more generally refers to the entire region downwind of the MD.

If the magnetic field is weak compared to HD effects, the astrospheric structure is rotationally symmetric about the axis that is parallel to the upwind and downwind directions and passes through the star, namely, the central axis. In HD models, this line is typically referred to as the stagnation line because an ISM flow along this axis would have to stagnate some distance after crossing the BS in order to preserve the rotational symmetry. In MHD a stagnation point, where the fluid comes to rest, may occur as well but generally does not lie on the central axis because flow lines are bent by the magnetic field \citep[e.g.][Sect.~2.1]{scherer+2020}.

Within this study the astrosphere around the runaway O-star \textlambda~Cephei has been used as a basis for the modelling procedure \citep[cf.][]{scherer+2020,baalmann+2020,baalmann+2021}. Following the classification scheme proposed by \citet{henney+2019a,henney+2019b,henney+2019c}, this astrosphere features a wind-supported BS, that is, the dynamics are determined by the MHD flows instead of the stellar radiation, and the interstellar dust is highly coupled to the plasma. This justifies the use of single-fluid non-radiative MHD to model the astrosphere, taking the stellar radiation and the interstellar dust into account only as energy gains and losses via heating and cooling effects, respectively (cf.~Sect.~\ref{sec:mhd}).

The structure of this paper is as follows. Section~\ref{sec:methods} gives a brief overview of the applied methodology, Sect.~\ref{sec:res} discusses the results, and Sect.~\ref{sec:summ} provides a summary and draws conclusions.

\section{Methodology}\label{sec:methods}

This section gives an overview of the methodology used in this investigation. The computational model is described in Sect.~\ref{sec:mhd} and details of this approach can be found in \citet{scherer+2020} and \citet[][Sect.~2.4]{baalmann_phd}. The projection method is summarised in Sect.~\ref{sec:proj}. It was first introduced in \citet{baalmann+2020} and expanded upon in \citet{baalmann+2021}. A detailed formulation is given in \citet[][Ch.~3]{baalmann_phd}. The setup of this investigation is described in Sect.~\ref{sec:setup}.

\subsection{Computational model}\label{sec:mhd}

The model astrospheres were computed as described by \citet[][Sect.~2.1]{baalmann+2021}, using the semi-discrete finite-volume code \textsc{Cronos} \citep{kissmann+2018} to solve the single-fluid ideal MHD equations:
\begin{align}
 \frac{\partial n}{\partial t} + \nabla \cdot \left( n \vec{u}\right) &= 0 \ ,\label{eq:mhdcont}\\
 \frac{\partial}{\partial t} \left( mn \vec{u} \right) + \nabla \cdot \left( m n \vec{u} \otimes \vec{u}\right) + \nabla p + \frac{1}{\mu_0} \vec{B} \times \left( \nabla \times \vec{B} \right) &= \vec{0} \ ,\label{eq:mhdimp}\\
 \frac{\partial e}{\partial t} + \nabla \cdot \left[ \left( e + p + \frac{1}{2\mu_0} \left| \vec{B} \right|^2 \right) \vec{u} - \frac{1}{\mu_0}\left( \vec{u} \cdot \vec{B} \right) \vec{B}\right] &= \Gamma-\Lambda \ ,\label{eq:mhdenergy}\\
 \frac{\partial \vec{B}}{\partial t} - \nabla \times \left( \vec{u} \times \vec{B} \right) &= \vec{0} \ ,\label{eq:mhdindu} 
\end{align}
using an HLL Riemann solver and a second-order Runge-Kutta scheme. In the above equations $n$, $\vec{u}$, $p$, $\vec{B}$, and $e$ are the number density, fluid velocity, thermal pressure, magnetic induction, and total energy density, respectively; $m$ and $\mu_0$ are the proton mass and vacuum permeability, respectively. The dyadic product is represented by $\otimes$. Two additional equations close the above system:
\begin{align}
 e &= \frac{p}{\gamma -1} + \frac{1}{2}mn \left| \vec{u} \right|^2 + \frac{1}{2\mu_0}\left| \vec{B} \right|^2 \ ,\label{eq:mhdgas}\\
 \nabla \cdot \vec{B} &= 0 \ ,
 \label{eq:energy_closure}
\end{align}
assuming a mono-atomic ideal gas, leading to the polytropic index $\gamma=5/3$. A source term was added to the right-hand side of the energy equation (\ref{eq:mhdenergy}), incorporating gains, $\Gamma$, and losses, $\Lambda$, of the total energy density by heating and cooling effects, respectively. Heating was incorporated as per \citet{kosinski+2006}, including photoionisation ($\propto\! n^2$), and photoelectric heating by dust, Coulomb collisions with cosmic rays, and the dissipation of interstellar turbulence ($\propto\! n$). Radiative cooling was incorporated following \citet{schure+2009} within the temperature range $\log(T\,[\si{K}])\in\left[3.8, 8.16\right]$; for lower temperatures, cooling effects were neglected; higher temperatures did not occur. As discussed in detail in \citet[][Sect.~3.1]{baalmann+2021}, the terms referred to as heating and cooling, which change the total energy density as per Eq.~(\ref{eq:mhdenergy}), have only an insignificant influence on the temperature of the plasma due to the additional energy input being continuously and self-consistently redistributed between the constituent terms of Eq.~(\ref{eq:energy_closure}). 

The computational grid takes the shape of a full sphere, described by spherical coordinates, $(r, \vartheta, \varphi)$, where $r\in[r_{\mathrm{SW}}, r_{\mathrm{ISM}}]$ is the radial coordinate, that is, the distance from the coordinate centre ($r=0$), which corresponds to the position of the star; $r_{\mathrm{SW}}>0$ and $r_{\mathrm{ISM}}\gg r_{\mathrm{SW}}$ are the inner and outer boundaries of the model (Table~\ref{tab:vismvals}), at which the respective boundary conditions (top segment of Table~\ref{tab:simvals}) for the SW and the ISM are set. More precisely, these are kept fixed for the inner ($r_{\mathrm{SW}}$) and the inflow-upwind half of the outer ($r_{\mathrm{ISM}}$) spherical boundary, whereas at the outflow-downwind half of the outer boundary all values are extrapolated linearly into the boundary cells. Additionally, no material is allowed to flow back into the computational domain through the downwind half-sphere by prohibiting any inward motion of material there, $u_r \ge 0$. The stellar parameters of \textlambda~Cephei are given in the bottom segment of Table~\ref{tab:simvals}. The angles $\vartheta\in[-90\si{\degree}, 90\si{\degree}]$ and $\varphi\in[-180\si{\degree}, 180\si{\degree}]$ are the polar and the azimuthal angle, respectively. Model cells are arranged with equidistant intervals in $r$ and equiangular intervals in $\vartheta,\varphi$; thus, the spatial extents of the cells perpendicular to $r$ as well as their volumes increase with $r$. The coordinates $(r_i, \vartheta_i, \varphi_i)$ of model cell $i$ describe the central point of that cell. The numbers of cells in the $\{r,\vartheta,\varphi\}$-direction are referred to as $N_{\{r,\vartheta,\varphi\}}$; the radial and angular extents of the cell sizes are referred to as $\Delta\{r,\vartheta,\varphi\}$ (Table~\ref{tab:vismvals}). 

\begin{table}
    \caption{\label{tab:vismvals}Radial grid properties of the models with different ISM inflow speeds in 3D (\textit{top segment}) and 2D (\textit{bottom segment}); cf.~\citet[][Table~4.2]{baalmann_phd}.}
    \centering
    \begin{tabular}{cccc}
        \toprule
        $u_{\mathrm{ISM}}\,[\si{km/s}]$ & $r_{\mathrm{SW}}\,[\si{pc}]$ & $r_{\mathrm{ISM}}\,[\si{pc}]$ & $\Delta r\,[\si{mpc}]$ \\ \midrule
        80 & 0.05 & \phantom{00}5 & \phantom{00}2.417 \\
        40 & 1\phantom{.00} & \phantom{0}20 & \phantom{0}18.55\phantom{0} \\
        20 & 1\phantom{.00} & \phantom{0}80 & \phantom{0}77.15\phantom{0} \\
        10 & 1\phantom{.00} & 150 & 145.5\phantom{00} \\ \midrule
        80 & 0.05 & \phantom{00}5 & \phantom{00}4.342 \\
        40 & 0.2\phantom{0} & \phantom{0}20 & \phantom{0}17.37\phantom{0} \\
        20 & 0.8\phantom{0} & \phantom{0}80 & \phantom{0}69.47\phantom{0} \\
        10 & 1.5\phantom{0} & 150 & 129.8\phantom{00} \\
        \bottomrule
    \end{tabular}
    \tablefoot{The 3D models have been simulated on grids with $N_r\times N_{\vartheta}\times N_{\varphi}=1024\times 64\times 128$\,cells, except for $u_{\mathrm{ISM}}=80\,\si{km/s}$, which has been modelled with $N_r=2048$\,cells; the angular numbers of cells correspond to the angular resolution $\Delta\vartheta=\Delta\varphi=2.8125\si{\degree}$. The 2D models use grids with $1140\times 240\times 1$\,cells, corresponding to $\Delta\vartheta=0.75\si{\degree}$.}
\end{table}

\begin{table}
    \caption{\label{tab:simvals}Boundary values corresponding to the astrospheres around the modelled stars (\textit{top segment}), and stellar parameters of \textlambda~Cephei (\textit{bottom segment}); cf.~\citet[][Table~2.3]{baalmann_phd}.}
    \centering
    \begin{tabular}{lrllr}
        \toprule
        \multicolumn{2}{l}{Variable} & Unit & \multicolumn{1}{c}{SW} & \multicolumn{1}{c}{ISM}  \\ \midrule
        Number density & $n$ & $[\si{cm^{-3}}]$ & $\phantom{000}3.4$ & $11$ \\
        Bulk speed & $u$ & $[\si{km\,s^{-1}}]$ & $2500$ & ($\ast$) \\ 
        Temperature & $T$ & $[10^3\,\si{K}]$ & $\phantom{00}10$ & $9$ \\
        Magn.\ induction & $B$ & $[\si{nT}]$ & $3\times 10^{-3}$ & $1$ \\
        Magn.\ polar angle & $\vartheta_B$ & $[\si{\degree}]$ &  & $30$ \\
        Magn.\ azim.\ angle & $\varphi_B$ & $[\si{\degree}]$ &  & $150$  \\ \midrule
        \multicolumn{3}{l}{Spectral type} & \multicolumn{2}{c}{O6If(n)p} \\
        Effect.\ temperature & $T_{\mathrm{eff}}$ & $[10^3\,\si{K}]$ & \multicolumn{2}{c}{$\phantom{0}36\phantom{.0}$} \\
        Distance (\textit{Gaia}) & $d_0$ & $[\si{pc}]$ & \multicolumn{2}{c}{$617\phantom{.0}$} \\
        Luminosity & $L_{\ast}$ & $[L_{\odot}]$ & \multicolumn{2}{c}{$6.3\times10^{5\phantom{-}}$} \\
        Mass & $M_{\ast}$ & $[M_{\odot}]$ & \multicolumn{2}{c}{$\phantom{0}51.4$} \\
        Rotation period & $P_{\ast}$ & $[\si{d}]$ & \multicolumn{2}{c}{$\phantom{00}4.1$} \\ 
        Mass loss rate & $\dot{M}_{\ast}$ & $[M_{\odot}\,\si{yr^{-1}}]$ & \multicolumn{2}{c}{$6.8\times 10^{-7}$} \\
        \bottomrule
    \end{tabular}
    \tablefoot{The boundary condition of $u_{\mathrm{ISM}}$ ($\ast$) has been set up as per Table~\ref{tab:vismvals}. Values taken from \citet{scherer+2016} and \citet[][Table~2]{scherer+2020}; cf.\ references therein; distance derived from \textit{Gaia} measurements \citep{gaia2018}.}
\end{table}

For the 3D models the cell sizes in both angular directions are set to be identical, $\Delta\vartheta=\Delta\varphi$, $2N_{\vartheta}=N_{\varphi}$. The angles of the homogeneous ISM fluid inflow are set to $\vartheta_u=90\si{\degree}$ and $\varphi_u=180\si{\degree}$, aligning the inflow with the $x$-axis. Thus, the $x$-axis points in the upwind direction, which is the direction from where the ISM inflow comes, or, conversely, in which the star moves. The direction in which the negative $x$-axis points is referred to as the downwind or tail direction. The SW is set up as an isotropic outflow. The homogeneous interstellar magnetic field is angled with $\vartheta_B$ and $\varphi_B$; the SW magnetic field is compliant with Parker's spiral \citep{parker1958} for the case of a purely radial, constant, monopolar field strength at the boundary surface, as implemented by \citet[][Sect.~3.1.1]{scherer+2020},
\begin{equation}
    \vec{B}_{\mathrm{SW}}(\vec{r})=B_{0}\frac{r_{\mathrm{SW}}^2}{r^2}\left(\vec{e}_r+\frac{r \Omega_{\mathrm{SW}}}{u_{\mathrm{SW}}}\sin\vartheta\vec{e}_{\varphi}\right) \ ,
\end{equation}
where $B_0$ is a prefactor chosen to comply with $\left|\vec{B}_{\mathrm{SW}}\right|_{r=r_\mathrm{SW}}=B_{\mathrm{SW}}$; $\vec{e}_{\{r,\varphi\}}$ are the respective unit vectors, and $\Omega_{\mathrm{SW}}=2\pi/P_{\ast}$ is the angular frequency of stellar rotation.

For the 2D models the $\varphi$-direction is covered by a single cell, $N_{\varphi}=1$, $\Delta\varphi=360\si{\degree}$, forcing cylindrical symmetry about the $z$-axis. The ISM inflow comes from the positive $z$-direction, $\vartheta_u=0\si{\degree}$. To maintain symmetry the interstellar magnetic field must be parallel to the $z$-axis, $\vartheta_B\in\{0\si{\degree},180\si{\degree}\}$, and the SW magnetic field must be zero, $\vec{B}_{\mathrm{SW}}=\vec{0}$.

The simulation times of all models are normalised to the time it takes to travel $1\,\si{pc}$ at the Alfv\'en speed in a medium with a number density of $n=1\,\si{cm^{-3}}$ and a magnetic field of $1\,\si{nT}$, which is $v_{\mathrm{A,0}}=1\,\si{nT}\cdot(1\,\si{cm^{-3}}\mu_0 m_{\mathrm{p}})^{-1/2}\approx 21.812\,\si{km/s}$, resulting in a time normalisation of
\begin{equation}
    t_{\mathrm{pc}}=\frac{1\,\si{pc}}{v_{\mathrm{A,0}}}\approx 44\,829\,\si{yr} \ .
\end{equation}

\subsection{Projection method}\label{sec:proj}

In order to generate synthetic observational images, the models were projected onto a virtual sky as described by \citet{baalmann+2020}. To this effect the model was shifted to a desired position, $(d_0, b_0, l_0)$, in Galactic coordinates, where $d_0$ is the distance modulus of the star, $b$ its Galactic latitude, and $l$ its Galactic longitude. For all projections in this work $b_0=l_0=0$ was chosen because these coordinates merely function as an offset of the image on the projection grid. The 2D models were extruded to cylindrically symmetric 3D models with $2N_{\vartheta}=N_{\varphi}$ before projection. Because the models have the shape of a sphere, they appear as a disk with a radius of
\begin{equation}
    \phi_{\max}=\arcsin\left(\frac{r_{\max}}{d_0}\right) \ ,
\end{equation}
where $r_{\max}=r_{\mathrm{ISM}}-\Delta r/2$ is the radius of the outermost model cell's centre. The largest spatial distance between two adjacent model cells is 
\begin{equation}
    (\Delta s)_{\max}=\max\left\{\Delta r,\ 2r_{\max}\sin\left(\frac{\Delta\vartheta}{2}\right)\right\} \ ,
\end{equation}
corresponding to the largest apparent distance between two model cells within the projection,
\begin{equation}
    (\Delta\phi)_{\max}=2\arctan\left(\frac{\frac{1}{2}(\Delta s)_{\max}}{d_0-r_{\max}}\right) \ .
\end{equation}
The square projection grid of size $\phi_{\max}\times\phi_{\max}$ is divided into $N\times N$ pixels, $N\in\{2,3\}$. The pixels are recursively refined by the same division until the pixel size would be smaller than $(\Delta\phi)_{\max}$. Because the pixel resolution, $\Delta\phi$, is limited by the spatial distance of the model's outermost cells, $(\Delta s)_{\max}$, due to the constant angular cell extent, $\Delta\vartheta$, it cannot adequately resolve the model's inner structure. Therefore, the model is linearly interpolated in both angular coordinates by a factor of $k=5$ before projection. 

The model cells are assigned to the pixel in which their Galactic coordinates appear, the projection value of each pixel is computed by summing the H\textalpha~radiance of all its assigned model cells,
\begin{equation}\label{eq:halpha}
    L_{\Omega,i}=\alpha_{\mathrm{eff},3\rightarrow 2}(T_i) \,h\nu_{\text{H\textalpha}} \,n_i^2 \frac{V_i}{4\pi d_i^2}\frac{1}{(\Delta\phi)^2} \ ,
\end{equation}
where $V_i$ and $d_i$ are the volume and distance of model cell $i$, $h$ is the Planck constant, $\nu_{\text{H\textalpha}}=456.81\,\si{THz}$ is the frequency of H\textalpha\ photons, and $\alpha_{\mathrm{eff},3\rightarrow 2}(T_i)$ is the temperature-dependent effective recombination rate coefficient, approximated by 
\begin{equation}
    \alpha_{\mathrm{eff},3\rightarrow 2}(T_i) \approx \alpha_3(T_i) + \sum\limits_{q=4}^{16} \alpha_q(T_i) \,P_{q,3} \ ,
\end{equation}
where $\alpha_q(T_i)$ are the temperature-dependent recombination rate coefficients of state $q$, taken from \citet{maokaastra2016}, and $P_{q,3}$ are the branching ratios from state $q$ through state $3$, computed from data by \citet{wiesefuhr2009} as described by \citet{baalmann+2020}. The highest radiance is expected to come from the outer astrosheath, where the density is highest and the temperature is comparably low \citep{baalmann+2020}. 

\subsection{Investigative setup}\label{sec:setup}

The objective of this examination was to compare otherwise identical astrospheric models with differing ISM inflow speeds, $u_{\mathrm{ISM}}$, including a model that would not generate a BS, in either 2D or 3D. As a prototype the astrosphere of the runaway O-star \textlambda~Cephei was chosen, cf.~\citet{baalmann+2020,baalmann+2021}. The stellar parameters of \textlambda~Cephei are given in the bottom segment of Table~\ref{tab:simvals}; the initial boundary conditions of the model are listed in the top segment of the same table. The relative speed between \textlambda~Cephei and its surrounding ISM is roughly $u_{\mathrm{ISM}}=80\,\si{km/s}$. The model astrosphere with this $u_{\mathrm{ISM}}$ is identical to the \texttt{hires} model presented by \citet{baalmann+2021}.

The sonic and Alfv\'enic speeds, $v_{\{\mathrm{c,A}\}}$, for the surrounding ISM are
\begin{align}
    v_{\mathrm{c,ISM}}&=\sqrt{\frac{\gamma p_{\mathrm{ISM}}}{\rho_{\mathrm{ISM}}}}\approx 15.74\,\si{km/s} \ , \\
    v_{\mathrm{A,ISM}}&=\frac{B_{\mathrm{ISM}}}{\sqrt{\mu_0 \rho_{\mathrm{ISM}}}}\approx 6.58\,\si{km/s} \ ,
\end{align}
where $p_{\mathrm{ISM}}=2n_{\mathrm{ISM}}k_{\mathrm{B}}T_{\mathrm{ISM}}$ is the thermal pressure of the ISM, $k_{\mathrm{B}}$ is the Boltzmann constant factor $2$ comes from the quasi-neutrality and pressure equilibrium of protons and electrons, while $\rho_{\mathrm{ISM}}=(m_{\mathrm{p}}+m_{\mathrm{e}})n_{\mathrm{ISM}}\approx m_{\mathrm{p}}n_{\mathrm{ISM}}$ is the mass density of the ISM. The characteristic speed for MHD shocks is the fast magnetosonic speed,
\begin{align}
    v_{\mathrm{f,ISM}}&=\left.\sqrt{\frac{1}{2}\left(v_{\mathrm{A}}^2+v_{\mathrm{c}}^2+\sqrt{\left(v_{\mathrm{A}}^2+v_{\mathrm{c}}^2\right)^2-4v_{\mathrm{A}}^2 v_{\mathrm{c}}^2 \cos^2\vartheta_{v,B}}\right)}\right|_{\mathrm{ISM}} \nonumber\\
    &\in\left[\max\left\{v_{\mathrm{c,ISM}},\ v_{\mathrm{A,ISM}}\right\},\ \sqrt{v_{\mathrm{c,ISM}}^2+v_{\mathrm{A,ISM}}^2}\right] \\
    &\approx \left[15.74,\ 17.06\right]\,\si{km/s} \ , \nonumber
\end{align}
which depends on the angle between the bulk velocity and the magnetic field, $\vartheta_{v,B}$.
The speed of the ISM inflow speed, $u_{\mathrm{ISM}}$, was incrementally decreased by a factor of two until it reached a subfast value, yielding $u_{\mathrm{ISM}}\in\{80, 40, 20, 10\}\,\si{km/s}$. Because the size of the shock structure is dependent on, for instance, the ISM inflow speed, models with smaller $u_{\mathrm{ISM}}$ require larger outer boundaries of the simulation grid. To keep computational times low, the inner boundaries were adapted similarly. The radii of the inner and outer boundaries, $r_{\{\mathrm{SW,ISM}\}}$, for the different ISM inflow speeds, as well as the radial resolutions, $\Delta r$, are given in Table~\ref{tab:vismvals}. 

\section{Results}\label{sec:res}

The results of the investigation are discussed in this section. The 2D models and the 3D models are analysed in Sects.~\ref{sec:2dmods}~\&~\ref{sec:3dmods}, respectively. In order to better understand the formation of the structures displayed in these sections, the evolution of the models is expounded at the example of a 2D model in Sect.~\ref{sec:evol}. Synthetic observations are presented in Sect.~\ref{sec:synthobs}. Further aspects are discussed by \citet[][Sect.~4.4]{baalmann_phd}.

\subsection{2D models}\label{sec:2dmods}

\begin{figure*}
    \centering
    \includegraphics[width=0.9\linewidth]{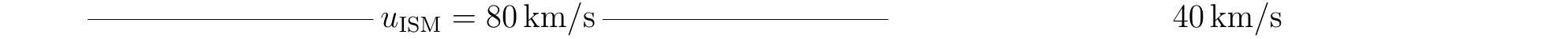}
    \includegraphics[width=0.9\linewidth]{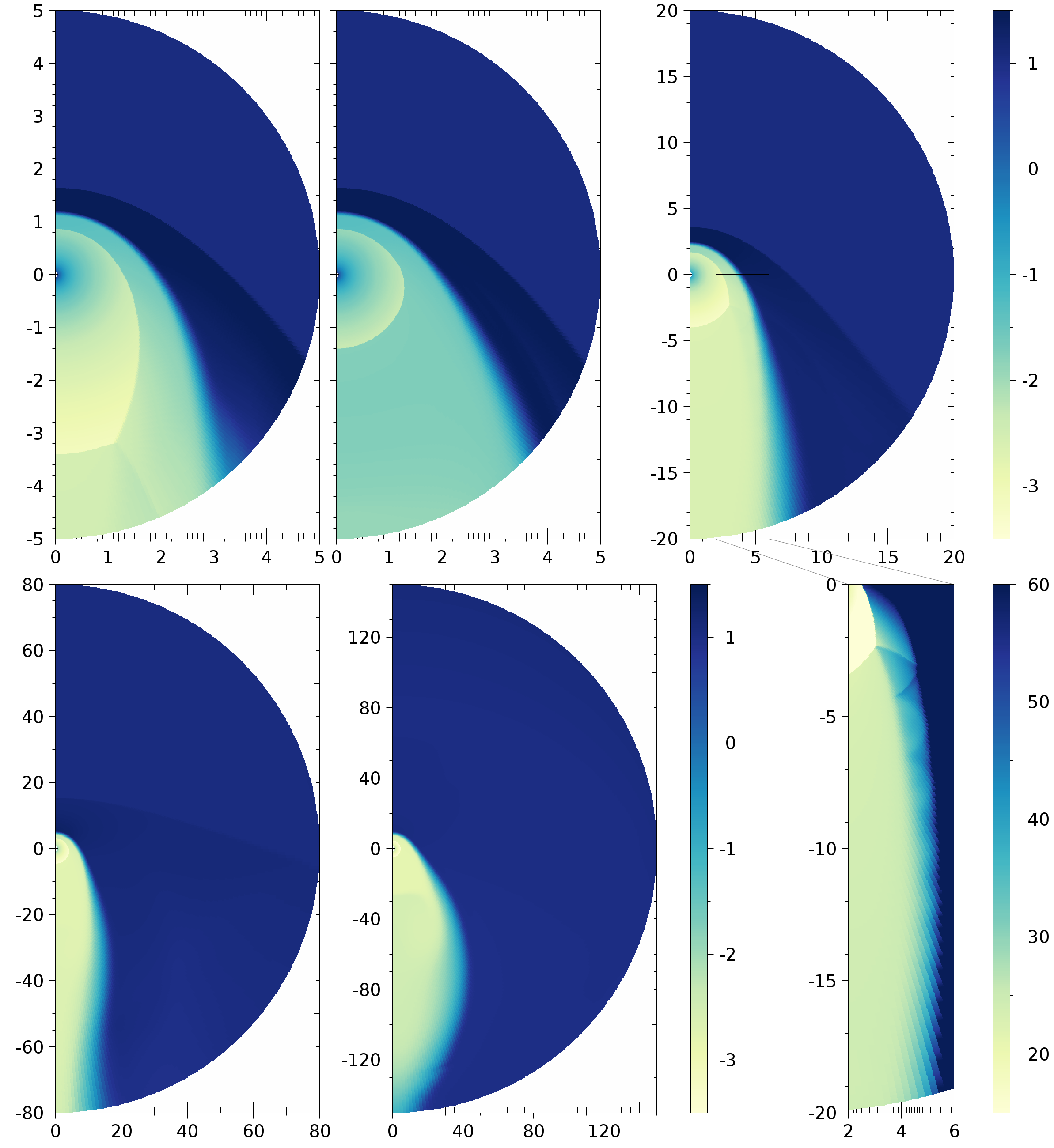}
    \includegraphics[width=0.9\linewidth]{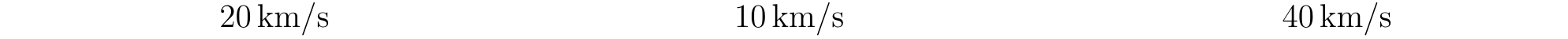}
    \caption{Slice planes of the 2D \textlambda-Cephei-like models at different ISM inflow speeds. From left to right and top to bottom: $u_{\mathrm{ISM}}=80\,\si{km/s}$ in its most distended state, $u_{\mathrm{ISM}}=80\,\si{km/s}$ in its most compressed state, $u_{\mathrm{ISM}}=40\,\si{km/s}$, $u_{\mathrm{ISM}}=20\,\si{km/s}$, $u_{\mathrm{ISM}}=10\,\si{km/s}$; number density, $\log(n)$, in $[\si{cm^{-3}}]$. Bottom right panel: close-up of the inner astrosheath for $u_{\mathrm{ISM}}=40\,\si{km/s}$; number density, $n$, in $[10^{-3}\,\si{cm^{-3}}]$. Axes in $[\si{pc}]$. Taken with permission from \citet[][Fig.~4.16]{baalmann_phd}.}
    \label{fig:lcep_vism_2d}
\end{figure*}

Slice planes of the 2D models are presented in Fig.~\ref{fig:lcep_vism_2d}. The vertical axes are the $z$-axis of the respective model; the ISM inflow comes from the top of the respective panel. Because 2D models are symmetric about the central \mbox{($z$-)}axis, the horizontal axes can be either the $x$- or $y$-axis or any other axis within the $xy$-plane. Figure~\ref{fig:lcep_vism_2d} shows contour plots in the logarithmic number density, $\log(n)$, measured in $[\si{cm^{-3}}]$. The bottom-right panel is a linearly rescaled closeup of the top-right panel. Due to their differences in size and ISM inflow speed, the models have taken different times to reach stationarity. The model with $u_{\mathrm{ISM}}=80\,\si{km/s}$ has reached stationarity of its upwind structure after $4t_{\mathrm{pc}}\approx179\,\si{kyr}$, however, its downwind structure vacillates (discussed below). The two top-left panels depict the astrosphere in its most distended (left panel) and most compressed (centre panel) state at $19.96t_{\mathrm{pc}}\approx895\,\si{kyr}$ and $18.96t_{\mathrm{pc}}\approx850\,\si{kyr}$, respectively. The model with $u_{\mathrm{ISM}}=40\,\si{km/s}$ (top right panel) is depicted at $30t_{\mathrm{pc}}\approx 1.3\,\si{Myr}$, although its structure within the innermost $8\,\si{pc}$ already reaches stationarity at $23t_{\mathrm{pc}}\approx 1.0\,\si{Myr}$. The model with $u_{\mathrm{ISM}}=20\,\si{km/s}$ (bottom left panel) has reached stationarity at $300t_{\mathrm{pc}}\approx 13.4\,\si{Myr}$; its inner structure has already reached stationary after $70t_{\mathrm{pc}}\approx3.1\,\si{Myr}$. The model with $u_{\mathrm{ISM}}=10\,\si{km/s}$ (bottom centre panel) has not reached stationarity after $400t_{\mathrm{pc}}\approx17.9\,\si{Myr}$, at which point the simulation was stopped, although the changes during the previous $150t_{\mathrm{pc}}\approx7\,\si{Myr}$ are minor \citep[][Sect.~4.4.1]{baalmann_phd}. Compared to the age of \textlambda~Cephei, which \citet{bouret+2012} estimate to be $5\,\si{Myr}$, and the time since the star has been accelerated to its high relative speed $2.5\,\si{Myr}$ ago according to \citet{gvaramadze+2011}, only the simulation times of the models with $u_{\mathrm{ISM}}\in\{80, 40\}\,\si{km/s}$ are reasonable. 

The homogeneous region of $n=11\,\si{cm^{-3}}$ at the top of the panels is the ISM. The neighbouring region of higher densities is the outer astrosheath; the border between the two delineates the BS. The distances of the BS, the AP, and the TS along the central axis, $z$, are tabulated in Table~\ref{tab:shockstruc}. The model with $u_{\mathrm{ISM}}=10\,\si{km/s}$ does not generate a BS because the ISM inflow is subfast. The strengths of the BSs also vary, which can be measured well with the compression ratio, $s$, the ratio of the number density behind the shock (i.e.\ in the outer astrosheath) to that in front of the shock (i.e.\ in the ISM). For the model with $u_{\mathrm{ISM}}=80\,\si{km/s}$ the compression ratio is highest, $s_{80}=3.64$. The model with $u_{\mathrm{ISM}}=40\,\si{km/s}$ has a notably weaker shock, $s_{40}=2.73$, and the shock of the model with $u_{\mathrm{ISM}}=20\,\si{km/s}$ is weaker still, $s_{20}=1.29$. Another clear difference of the BS of the model with $u_{\mathrm{ISM}}=20\,\si{km/s}$ as compared to the other two BSs is the opening angle. Whereas the BSs of the models with $u_{\mathrm{ISM}}\in\{80,40\}\,\si{km/s}$ are similar in shape, featuring an opening angle of roughly $\theta_{\{80,40\}}\approx80\si{\degree}$, the BS opening angle of the model with $u_{\mathrm{ISM}}=20\,\si{km/s}$ is almost $\theta_{20}\approx150\si{\degree}$ \citep[][Sect.~4.4.1]{baalmann_phd}.

\begin{table}
    \caption{\label{tab:shockstruc}Stellar-centric distances of the BS, AP, TS, and the MD in its most compressed ($\min$) and distended ($\max$) positions along the central axis for the 2D and 3D models with $u_{\mathrm{ISM}}\in\{80,40,20,10\}\,\si{km/s}$.}
    \centering
    \begin{tabular}{cccccccc}
        \toprule
        $u_{\mathrm{ISM}}$ & \multicolumn{2}{c}{80} & \multicolumn{2}{c}{40} & \multicolumn{1}{c}{20} & \multicolumn{2}{c}{10} \\ \cmidrule(l{0.2em}){2-3}\cmidrule(l{0.2em}){4-5}\cmidrule(l{0.2em}){7-8}
        $[\si{km/s}]$ & 2D & 3D & 2D & 3D & 2D & 2D & 3D \\ \midrule
        BS & 1.64 & 1.73 & 3.63 & 4.06 & 15.4  & \phantom{$\geq\,$}-- & --\\
        AP & 1.2\phantom{0} & 1.23 & 2.5\phantom{0} & 2.44 & 5.3\phantom{0} & \phantom{$\geq\,$}9.6 & 6\phantom{.0} \\
        TS & 0.87 & 0.86 & 1.70 & 1.64 & 3.3\phantom{0} & \phantom{$\geq\,$}4.5 & 3.8\\
        MD$_{\min}$ & 1.40 & 1.80 & 3.87 & 3.05 & 4.5\phantom{0} & \phantom{$\geq\,$}4.5 & 5.0\\
        MD$_{\max}$ & 3.40 & 1.80 & 4.37 & 3.05 & 5.6\phantom{0} & $>\,$4.8 & 5.0 \\
        \bottomrule
    \end{tabular}
    \tablefoot{The 3D model with $u_{\mathrm{ISM}}=20\,\si{km/s}$ has not yet reached stationarity. The models with $u_{\mathrm{ISM}}=10\,\si{km/s}$ do not generate a BS. Cf.~\citet[][Sect.~4.4.1]{baalmann_phd}.}
\end{table}

All three superfast models feature a vacillating MD. This has been closely examined for the model with $u_{\mathrm{ISM}}=80\,\si{km/s}$, where the MD vacillates between distances along the downwind $z$-axis of $z_{\mathrm{MD,80,min}}=-1.4\,\si{pc}$ and $z_{\mathrm{MD,80,max}}=-3.4\,\si{pc}$ (cf.~Table~\ref{tab:shockstruc}). This periodic behaviour repeats every $2.70t_{\mathrm{pc}}\approx121\,\si{kyr}$; the phase of distension takes $1.00t_{\mathrm{pc}}\approx44.8\,\si{kyr}$ and the phase of compression accordingly $1.70t_{\mathrm{pc}}\approx79.2\,\si{kyr}$ \citep[][Sect.~4.4.1]{baalmann_phd}. 

At the time of compression (top left panel of Fig.~\ref{fig:lcep_vism_2d}), there is a visible corner between the TS and the MD, the TP, which is located at $(x,z)=(1.16\,\si{pc}, -3.18\,\si{pc})$. From the TP, the TD extends downwind; it separates the astrotail with a number density of $n\approx0.003\,\si{cm^{-3}}$ from the inner astrosheath with $n\approx0.005\,\si{cm^{-3}}$. A reflected shock (RS) also begins at the TP at an angle of $30\si{\degree}$ to the vertical; across this TP, the number density jumps from $n\approx0.004\,\si{cm^{-3}}$ at the side closer to the AP to $n\approx0.007\,\si{cm^{-3}}$ at the side closer to the TD \citep[][Sect.~4.1]{baalmann+2021}. At the time of distension (top centre panel), no such corner between the TS and the MD is visible in the number density, and neither are the TD or the RS. In other parameters, such as the fast magnetosonic Mach number, $M_{\mathrm{f}}=u/v_{\mathrm{f}}$, or the sonic Mach number, $M_{\mathrm{c}}=u/v_{\mathrm{c}}$, both the TP and the TD are visible \citep[][Sect.~4.4.1]{baalmann_phd}.

The MDs of the models with $u_{\mathrm{ISM}}\in\{40,20\}\,\si{km/s}$ vacillate as well, although their periodicity has not been examined more closely due to time constraints. However, for the model with $u_{\mathrm{ISM}}=40\,\si{km/s}$ the TP and TD are always visible in the number density, even at the astrosphere's most compressed state. The MD is located along the $z$-axis at $z_{\mathrm{MD,40,max}}=-4.37\,\si{pc}$ at its most distended state after $33t_{\mathrm{pc}}\approx1.48\,\si{Myr}$ and at $z_{\mathrm{MD,40,min}}=-3.87\,\si{pc}$ when most compressed after $37.8t_{\mathrm{pc}}\approx1.69\,\si{Myr}$ (cf.~Table~\ref{tab:shockstruc}). A RS extends from the TP and is reflected multiple times by the AP and the TD, as can be seen in the bottom right panel of Fig.~\ref{fig:lcep_vism_2d}. For the model with $u_{\mathrm{ISM}}=20\,\si{km/s}$, the astrosphere is most distended at $265t_{\mathrm{pc}}\approx 11.9\,\si{Myr}$ with $z_{\mathrm{MD,20,max}}=-5.6\,\si{pc}$ and most compressed at $288t_{\mathrm{pc}}\approx 12.9\,\si{Myr}$ with $z_{\mathrm{MD,20,min}}=-4.5\,\si{pc}$ (cf.~Table~\ref{tab:shockstruc}). The model with $u_{\mathrm{ISM}}=10\,\si{km/s}$ is in its most compressed state with $z_{\mathrm{MD,10,min}}=-4.5\,\si{pc}$ for the first time at $282t_{\mathrm{pc}}\approx 12.6\,\si{Myr}$, and has not stopped its subsequent distension when the simulation was terminated at $400t_{\mathrm{pc}}$; it is therefore unknown whether its structure vacillates or not. 

\begin{figure}
    \centering
    \includegraphics[width=\linewidth]{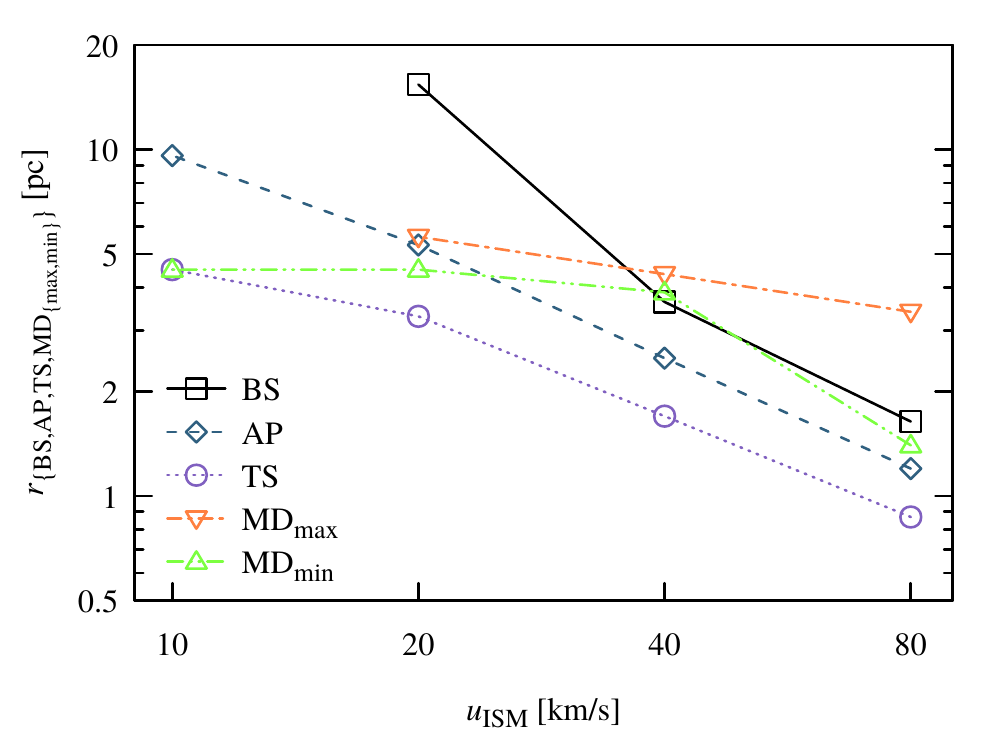}
    \caption{Distances, $r_{\{\mathrm{BS,AP,TS,MD_{\{\max,\min\}}}\}}$, of the BS (black squares, solid line), AP (blue diamonds, dashed line), TS (purple circles, dotted line), and the MD in its most distended (MD$_{\max}$, orange triangles, dash-dotted line) and compressed (MD$_{\min}$, green pyramids, dash-dot-dotted line) positions along the central axis of the 2D models (Table~\ref{tab:shockstruc}) against the respective inflow speed, $u_{\mathrm{ISM}}$. Both axes are scaled logarithmically.}
    \label{fig:dists}
\end{figure}

With the aim of probing the cause of the vacillating astrotails, a 2D model with $u_{\mathrm{ISM}}=80\,\si{km/s}$ and the resolution of the respective 3D model was generated. The astrotail of this model did not vacillate, indicating that the stronger influence of numeric diffusion on more coarsely resolved grids inhibits the vacillating behaviour.

In Fig.~\ref{fig:dists}, the stellar-centric distance of the discontinuities along the inflow axis of the 2D models as tabulated in Table~\ref{tab:shockstruc} are plotted against the respective inflow speed. Because the model with $u_{\mathrm{ISM}}=10\,\si{km/s}$ has not yet reached its most distended state, the distance of the distended MD is not plotted; there also is no BS for this model. In the dual logarithmic graph the distances of the TS, AP, and distended MD (MD$_{\max}$) lie on straight lines for superfast inflows, namely, for $u_{\mathrm{ISM}}\in\{20,40,80\}\,\si{km/s}$ with slopes of $-0.96$, $-1.08$, and $-0.36$, respectively. The respective distances of the subfast model, $u_{\mathrm{ISM}}=10\,\si{km/s}$, do not lie on these lines. Similarly, no linear behaviour on the dual logarithmic graph can be observed for the distances of the BS and the compressed MD (MD$_{\min}$). Unfortunately, the low number of available models does not allow to assign a high level of confidence to these results. 

\begin{figure}
    \centering
    \includegraphics[width=\linewidth]{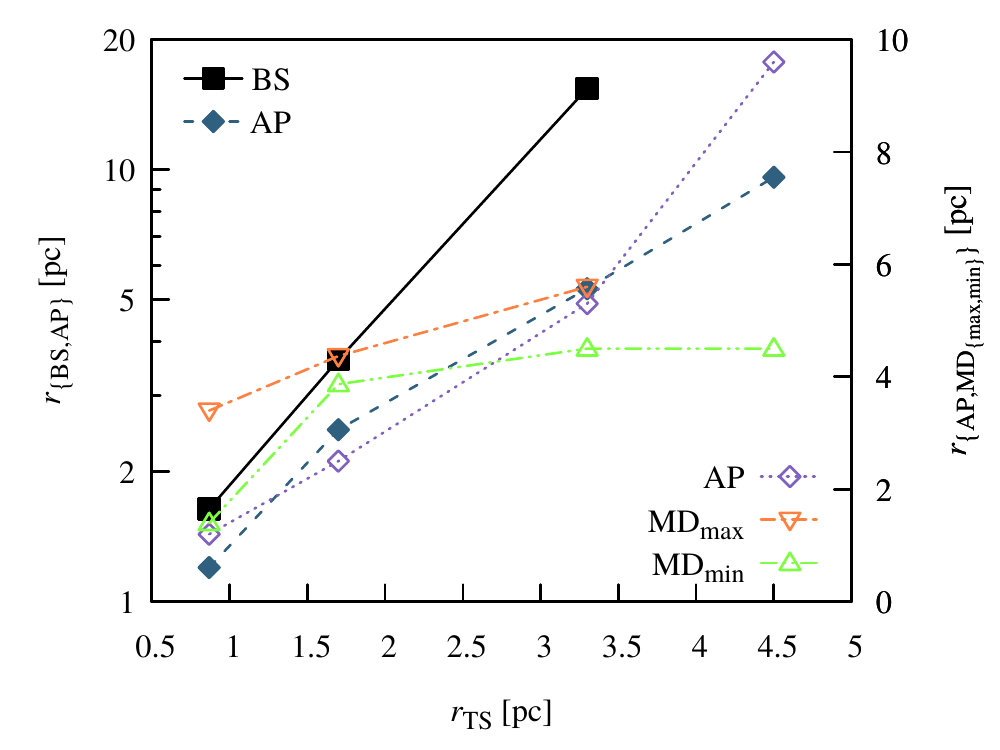}
    \caption{Distances, $r_{\{\mathrm{BS,AP,MD_{\{\max,\min\}}}\}}$, of the BS, AP, and the MD in its most distended ($\max$) and compressed ($\min$) positions along the central axis of the 2D models (Table~\ref{tab:shockstruc}) against the respective TS distance, $r_{\mathrm{TS}}$, making use of both logarithmic (left ordinate axis, filled symbols) and linear (right ordinate axis, empty symbols) scaling; the abscissa has been scaled linearly. The AP has been plotted on both scales. Colours as in Fig.~\ref{fig:dists} with the exception of the linearly scaled AP (purple dotted line).}
    \label{fig:dists2}
\end{figure}

In Fig.~\ref{fig:dists2}, stellar-centric distances of the BS, AP, and MD at both its most compressed and most distended states are plotted against the distance of the TS \citep[cf.][Fig.~6]{mueller+2006}. The ordinate axis for the distances of the BS and AP is scaled logarithmically, the ordinate axis for the two MD distances as well as the abscissa axis for the TS are scaled linearly. The distances of the BS appear to lie on a line on the semi-logarithmic axis system, as do the distances of the AP excluding the point for $u_{\mathrm{ISM}}=80\,\si{km/s}$ (leftmost in the graph). The distances of the AP excluding the point for $u_{\mathrm{ISM}}=10\,\si{km/s}$ (rightmost in the graph) also appear to lie on a line on the linear axis system, agreeing with Fig.~\ref{fig:dists}. The distances of the MD appear concave on the linear axis system. This only partially agrees with the findings of \citet{mueller+2006} based on a parameter study of 27 heliosphere-like models, who found linear correlations between the distances of the BS and AP to that of the TS. 

\subsection{2D evolution}\label{sec:evol}

\begin{figure*}
    \centering
    \includegraphics[width=0.9\linewidth]{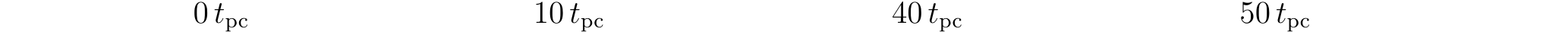}
    \includegraphics[width=0.9\linewidth]{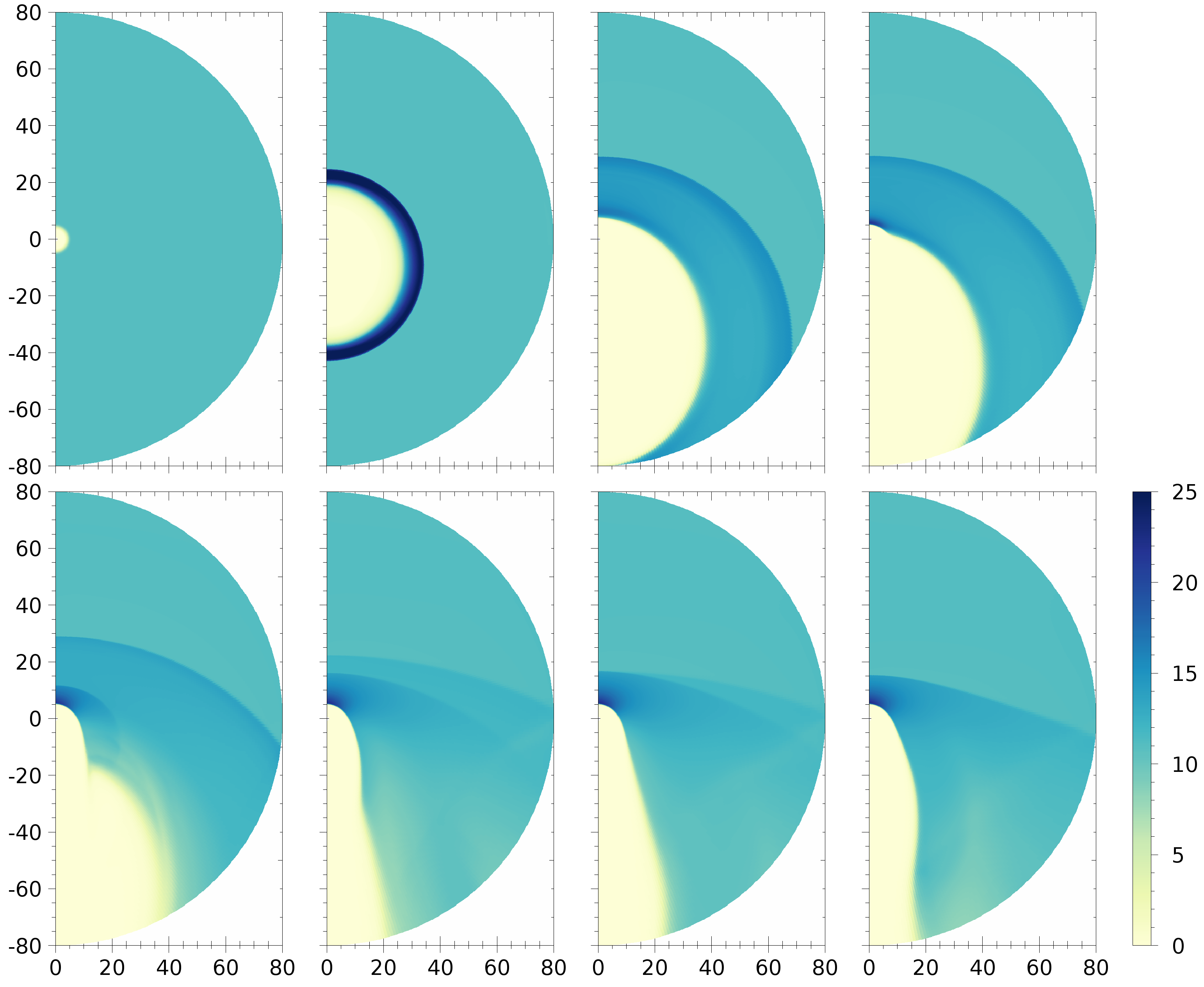}
    \includegraphics[width=0.9\linewidth]{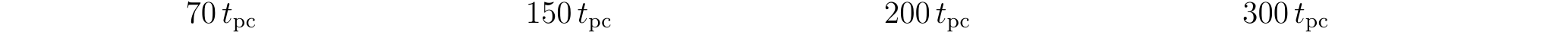}
    \caption{Slice planes of the 2D \textlambda-Cephei-like models with an ISM inflow speed of $u_{\mathrm{ISM}}=20\,\si{km/s}$ at times $t\in\{0,10,40,50,70,150,200,300\}\,t_{\mathrm{pc}}$ (from left to right and top to bottom) in the number density, $n$, in $[\si{cm^{-3}}]$; axes in $[\si{pc}]$. Taken with permission from \citet[][Fig.~4.17]{baalmann_phd}.}
    \label{fig:evol}
\end{figure*}

To facilitate the comprehension of the structures seen in the slice planes, the time evolution of the 2D model at $u_{\mathrm{ISM}}=20\,\si{km/s}$ from its inception until reaching stationarity has been analysed. This is depicted for selected times in Fig.~\ref{fig:evol}; the number density, $n$, is plotted linearly in order to better visualise the developing outer structure; the inner astrosphere cannot be resolved by the colour scale. At $t=0$ (top left panel), the model shows the initial conditions: the ISM as a homogeneous domain of $n=11\,\si{cm^{-3}}$ surrounding the radially decreasing gradient of the SW with $n<0.1\,\si{cm^{-3}}$, which cannot be resolved by the linear colour scale. The levelling between these two domains is continuous and smooth. The BS and AP move outwards until $t=10t_{\mathrm{pc}}\approx 448\,\si{kyr}$ (top centre-left panel); the MD and TS are unresolved. While these two inner shocks are still a single, perfectly spherical structure, the BS and AP are already distorted in the direction of the ISM inflow. After $t=10t_{\mathrm{pc}}$ the BS continues moving outwards, whereas the AP is being pushed downwind by the ISM. This can be seen at the depicted slice plane for $t=40t_{\mathrm{pc}}\approx 1.79\,\si{Myr}$ (top centre-right panel), where the BS has already moved out of the simulation grid in the downwind direction, the outer astrosheath has drastically grown in size, and plasma has piled up in front of the AP in the upwind direction. The ISM continues pushing the AP downwind but is balanced by the isotropic SW in the upwind direction. This causes a bulge on the AP close to the central axis, as can be seen at $t=50t_{\mathrm{pc}}\approx 2.24\,\si{Myr}$ (top right panel); more material piles up in front of it. At $t=70t_{\mathrm{pc}}\approx 3.14\,\si{Myr}$ (bottom left panel) the balance between the ISM and the SW has given the AP its bullet-like shape down to $z\approx-20\,\si{pc}$. A secondary BS moves from the AP outwards, whereas the original BS now moves inwards, cf.\ $t=150t_{\mathrm{pc}}\approx 6.72\,\si{Myr}$ (bottom centre-left panel). At $t=200t_{\mathrm{pc}}\approx 8.97\,\si{Myr}$ (bottom centre-right panel) the two BSs meet close to the central axis. The inner BS stays in position but the outer BS, farther away from the central axis, where it has not yet met the inner BS, moves further inwards. The AP has long since obtained its bullet-like shape within the entire simulation grid. At $t=300t_{\mathrm{pc}}\approx 13.4\,\si{Myr}$ (bottom right panel), both BSs have aligned; cf.\ the bottom left panel of Fig.~\ref{fig:lcep_vism_2d} for a logarithmic scaling at this time \citep[][Sect.~4.4.2]{baalmann_phd}.

This time evolution is most emphatically not compliant with the physical evolution of an astrosphere but is presented to qualitatively illustrate the dynamical behaviour of the system and the timescales on which it occurs. It also serves as an aid to understand the origin of the visible structures, which will be used in Sect.~\ref{sec:3dmods}. Similar behaviour can be observed for the models with $u_{\mathrm{ISM}}\in\{80,40\}\,\si{km/s}$, although their narrower outer astrosheaths drastically reduce the lifetime of the secondary BS. 

\subsection{3D models}\label{sec:3dmods}

\begin{figure*}
    \centering
    \begin{adjustbox}{width=0.9\linewidth}
    \includegraphics[height=100cm]{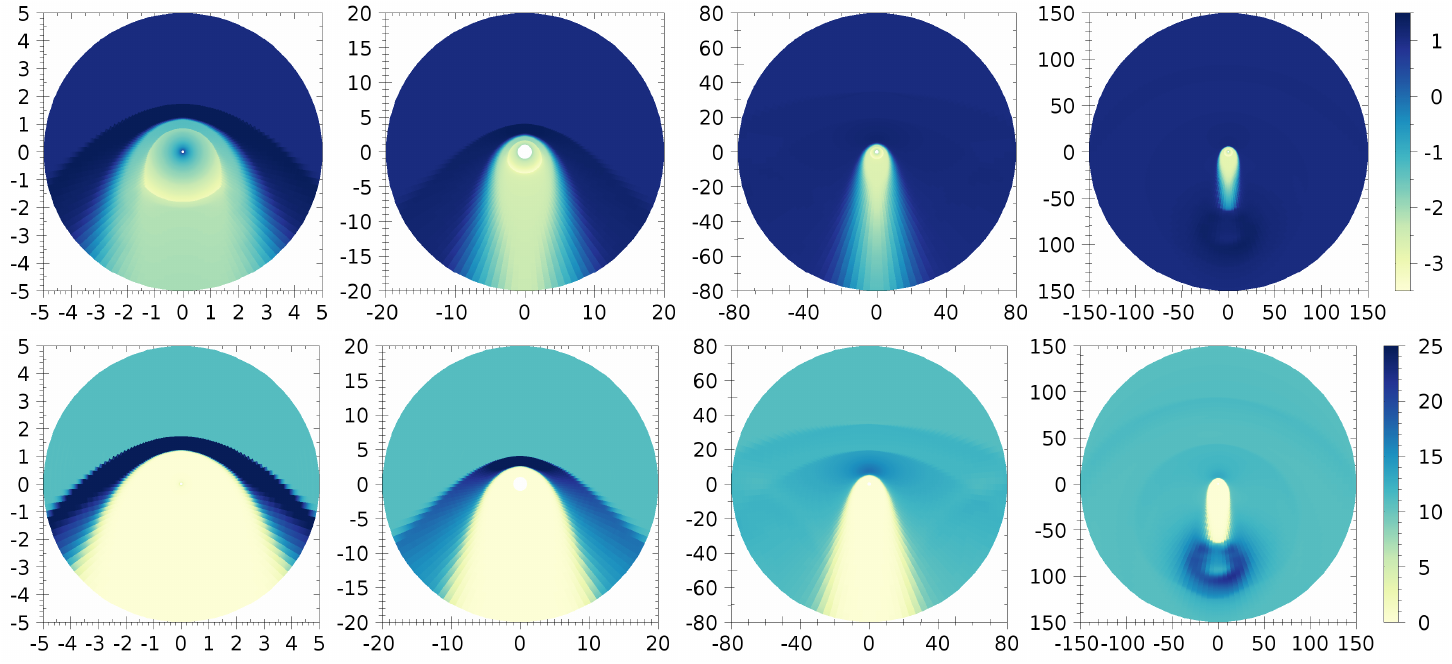}
    \includegraphics[height=100cm]{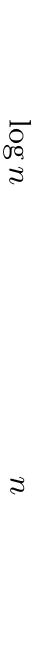}
    \end{adjustbox}
    \includegraphics[width=0.9\linewidth]{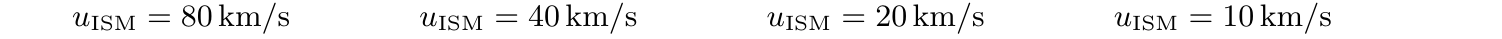}
    \caption{Slice of the $yx$-planes of the 3D \textlambda-Cephei-like models at different ISM inflow speeds, $u_{\mathrm{ISM}}\in\{80, 40, 20, 10\}\,\si{km/s}$ (\textit{from left to right}) in the number density, $\log(n)$ (\textit{top row}) and $n$ (\textit{bottom row}); number density in $[\si{cm^{-3}}]$, axes in $[\si{pc}]$. The two left images in the bottom row feature number densities in the outer astrosheath above the values covered by the colour scale, up to $n=41.5\,\si{cm^{-3}}$ and $n=30.6\,\si{cm^{-3}}$, respectively. Taken with permission from \citet[][Fig.~4.18]{baalmann_phd}.}
    \label{fig:lcep_vism_3d}
\end{figure*}

Cut slices of the corresponding 3D models are presented in Fig.~\ref{fig:lcep_vism_3d}, displaying the $yx$-plane ($y$-axis pointing rightwards, $x$-axis upwards) of the respective grid in the logarithmic number density, $\log(n)$, in the top row, and the linear number density, $n$, in the bottom row. As before, the linear colour scale cannot resolve the inner astrosphere (i.e.\ the TS and MD).

The 3D model with $u_{\mathrm{ISM}}=80\,\si{km/s}$ (leftmost column), depicted at its time of stationarity, $t=5.84t_{\mathrm{pc}}\approx 262\,\si{kyr}$, agrees well with its 2D counterpart. Its MD does not vacillate due to the coarser angular resolution, which increases the effects of numeric diffusion and therefore inhibits instabilities. The positions of the discontinuities along the central axis, cf.~Table~\ref{tab:shockstruc}, match those of the 2D model closely. Due to the comparatively coarse angular resolution, no RS is visible \citep[][Sect.~4.4.3]{baalmann_phd}. 

The 3D model with $u_{\mathrm{ISM}}=40\,\si{km/s}$ (centre-left column) shows only minor differences with regard to its 2D counterpart as well. It is depicted at its time of stationarity, $t=35t_{\mathrm{pc}}$, which is similar to that of the 2D model. The comparatively diffuse AP and the lack of a RS are artifacts of the coarse angular resolution. The positions of the discontinuities along the central axis are given in Table~\ref{tab:shockstruc} and agree fairly well with the 2D simulation, though it is notable that the MD lies further inward than its 2D counterpart even in its compressed state, and that the BS has a slightly larger opening angle \citep[][Sect.~4.4.3]{baalmann_phd}. 

The 3D model with $u_{\mathrm{ISM}}=20\,\si{km/s}$ (centre-right column) is depicted at $t=160t_{\mathrm{pc}}\approx7.17\,\si{Myr}$ and not yet stationary. Its shows a similar double-BS structure than its 2D counterpart \citep[][Sect.~4.4.3]{baalmann_phd}.

The 3D model with $u_{\mathrm{ISM}}=10\,\si{km/s}$ (rightmost column), however, shows significant differences with regard to its 2D counterpart. It is depicted at $180t_{\mathrm{pc}}\approx8.07\,\si{Myr}$, where the inner astrosphere is already stationarity but two bow waves still move outwards, akin to the two BSs of the previously examined models. The upwind structure of the 3D model lies considerably farther inwards compared to the 2D model, $x_{\mathrm{AP,10}}=6\,\si{pc}$ and $x_{\mathrm{TS,10}}=3.8\,\si{pc}$, compared to the 2D model's $z_{\mathrm{AP,10}}=9.6\,\si{pc}$ and $z_{\mathrm{TS,10}}=4.5\,\si{pc}$, whereas the MD is at a similar distance along the central axis, $x_{\mathrm{MD,10}}=-5\,\si{pc}$, compared to $z_{\mathrm{MD,10}}=-4.8\,\si{pc}$ (cf.~Table~\ref{tab:shockstruc}). The entire astrotail is contained by the simulation grid, which is in stark contrast to the 2D analogue. An additional structure downwind of the astrotail is visible for $x<-63.5\,\si{pc}$; the gradient of the number density significantly steepens its increase from $n<1\,\si{cm^{-3}}$ at $x>-62\,\si{pc}$ to $n>15\,\si{cm^{-3}}$ at $x<-68\,\si{pc}$. A structure of high density, $n>n_{\mathrm{ISM}}=11\,\si{cm^{-3}}$, lies farther downwind and continuously moves outwards. The material of this structure has been gathered by the second bow wave from the outer surface of the AP during the process of the AP's tightening, cf.\ the time evolution during $t\in[50, 150]t_{\mathrm{pc}}$ of the 2D model with $u_{\mathrm{ISM}}=20\,\si{km/s}$ (Sect.~\ref{sec:evol}). This structure is unstable, both moving downwind and dispersing at the depicted time. Because this structure's origin lies in the processes of the initial formation of the model astrosphere, which does not reflect the physical formation of a genuine astrosphere, the structure likely has no physical counterpart. Conceivably, a similar structure could be formed through different evolutionary phases of the SW and ISM. This structure does not exist in the 2D model; it is caused by the AP's severe asymmetry before its compression due to the second bow wave, which in turn is an effect of the oblique magnetic field \citep[][Sect.~4.4.3]{baalmann_phd}.

Due to the unphysical origin of this structure and the unrealistically high age of the astrosphere at its inception, a systematic study of this structure is unlikely to yield insightful results. However, stars with much longer lifetimes like M-type or even Sun-like stars may generate similar structures. It may be possible that such structures are common in astrospheric simulations but were not previously observed because most modelling approaches do not include the full astrotail (e.g. \citealp{katushkina+2018,meyer+2020,scherer+2020,herbst+2020,baalmann+2020}, but conversely \citealp{meyer+2021}).

\subsection{Synthetic observations}\label{sec:synthobs}

\begin{figure*}
    \centering
    \begin{adjustbox}{width=0.9\linewidth}
        \includegraphics[height=100cm]{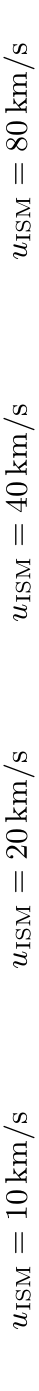}
        \includegraphics[height=100cm]{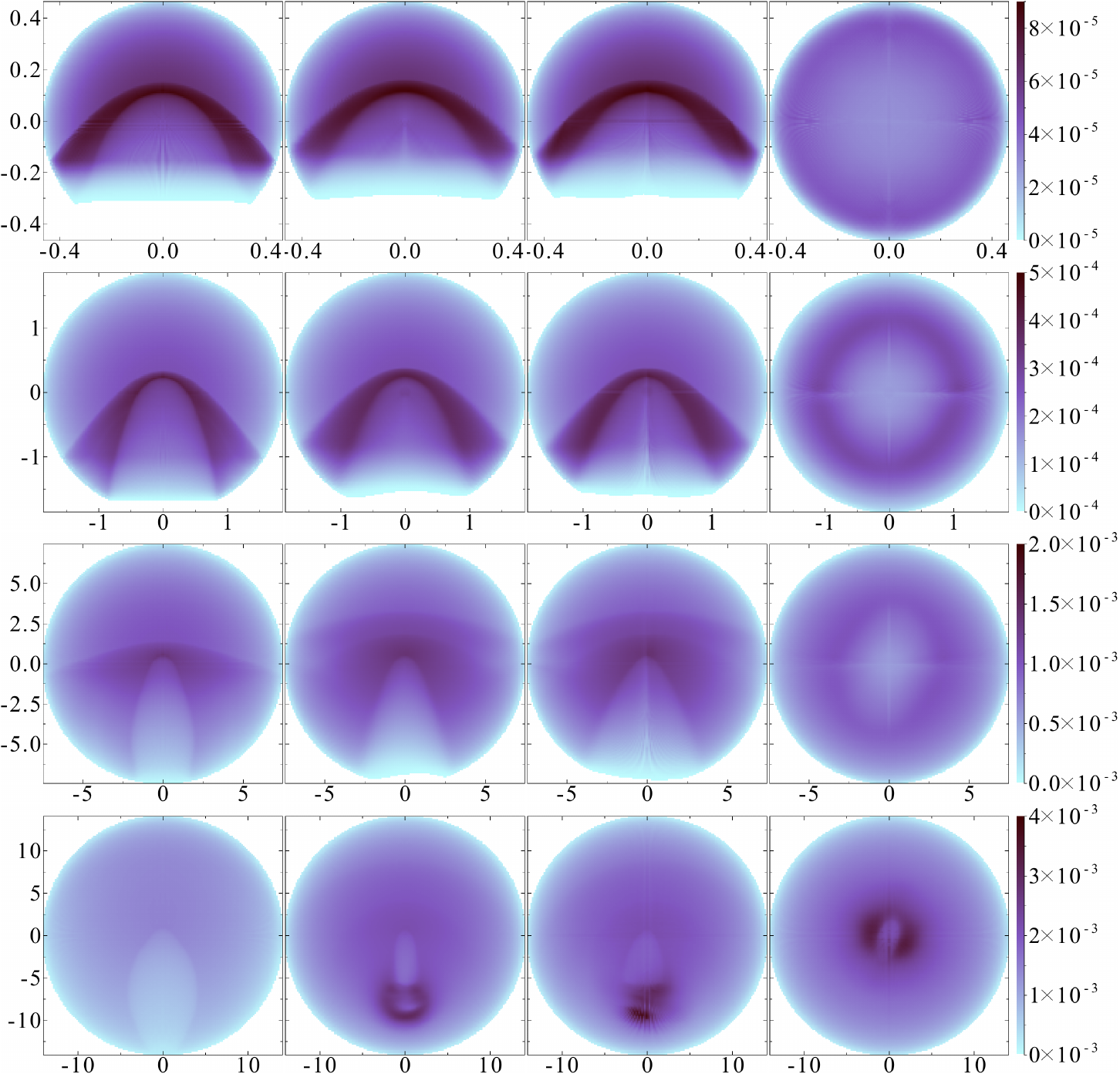}
    \end{adjustbox}
    \includegraphics[width=0.9\linewidth]{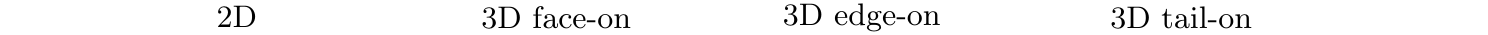}
    \caption{H\textalpha~projections of \textlambda-Cephei-like astrospheres with ISM inflow speeds (from top to bottom) $u_{\mathrm{ISM}}\in\{80, 40, 20, 10\}\,\si{km/s}$ onto the $lb$-plane; modelled in 2D (left), and in 3D projected face-on to the ecliptic (centre-left), edge on to the ecliptic (centre-right), and tail-on to the astrosphere (right). H\textalpha~radiance in $[\si{erg\,s^{-1}\,cm^{-2}\,sr^{-1}}]$, angular extent in degrees. Taken with permission from \citet[][Fig.~4.19]{baalmann_phd}.}
    \label{fig:vism_proj}
\end{figure*}

The different astrospheric models are projected in Fig.~\ref{fig:vism_proj} in the H\textalpha\ radiance as per Eq.~(\ref{eq:halpha}), at a stellar distance of $d_0=617\,\si{pc}$ \citep{gaia2018}. The four lines in the panels correspond to the four ISM inflow speeds, $u_{\mathrm{ISM}}\in\{80, 40, 20, 10\}\,\si{km/s}$ (from top to bottom). The first column displays synthetic observations of the 2D models, which have been cyclindrically extruded to 3D models before line-of-sight-integrating; the geometry of the projection has been set up to view the slice planes of Fig.~\ref{fig:lcep_vism_2d}, namely, the $yz$-plane of the extruded model, face-on. The three right columns show synthetic observations of the respective 3D models, in the centre-left column face-on to the slice plane of Fig.~\ref{fig:lcep_vism_3d}, which is the $yx$-plane, namely, the ecliptic, of the model; in the centre-right column the models are projected edge-on to the slice plane, namely, face-on to the $zx$-plane, and in the right column tail-on to the astrosphere, namely, face-on to the $zy$-plane. All projections are subject to Moir\'e patterns, which are visible, for example, as lines along the coordinate axes.

For $u_{\mathrm{ISM}}=80\,\si{km/s}$ (top row), the 2D projection and the face-on and edge-on projections of the 3D model agree well. This is as expected; because hydrodynamic effects are dominant over the magnetic field, as measured, for example, by the thermal and ram pressures versus the magnetic pressure, the 3D model must be similar to the 2D model and furthermore be reasonably symmetric about the central axis. Nevertheless, the opening angle of the 3D model's BS is slightly larger due to the added magnetic pressure. The tail-on projection is rotationally symmetric about its centre; its radiance is highest within an outer ring with a radius of about $0.4\si{\degree}$, where the lengths of the local lines of sight through the high-radiance outer astrosheath are longest \citep[][Sect.~4.4.4]{baalmann_phd}. 

A similar behaviour is apparent for $u_{\mathrm{ISM}}=40\,\si{km/s}$ (second row). Again, the opening angle of the 3D model's BS is slightly larger than its 2D counterpart. The ring of high radiance in the tail-on projection has a radius of about $1\si{\degree}$, corresponding to the lines of sight with the longest components within the outer astrosheath \citep[][Sect.~4.4.4]{baalmann_phd}. 

The rotational symmetry is broken for $u_{\mathrm{ISM}}=20\,\si{km/s}$ (third row), as the tail-on projection reveals. The ring of high radiance is no longer rotationally symmetric but elliptical. Its long axis is tilted by about $30\si{\degree}$ to the vertical axis. The face-on and edge-on projections are accordingly dissimilar; the opening angle of the edge-on view's BS is larger than its face-on counterpart. The difference with regard to the 2D projection comes mostly from the difference in astrospheric evolution: the 2D model is already stationary, whereas the 3D model is still evolving, as can be seen, for example, by the double-BS structure that is apparent in both the face-on and the edge-on projections \citep[][Sect.~4.4.4]{baalmann_phd}. 

Some of the dissimilarities between the 2D and the 3D model with $u_{\mathrm{ISM}}=10\,\si{km/s}$ (bottom row) are caused by the different times of projection as well. In the face-on and edge-on projections of the 3D model, the outbound bow waves are vaguely perceptible in the upwind direction, $x>0$, as disks of slightly higher radiance with radii of about $4\si{\degree}$ and $8\si{\degree}$, respectively; in the 2D projection, where the bow waves have already dispersed, the radiance within these regions is notably lower. The different shapes of the inner astrosphere, that is, those of the AP and the astrotail, are not caused by the difference in time, however. While the 2D structure is drop-shaped and extends out of the simulation grid in the downwind direction, the 3D structure varies in shape in the face-on and edge-on view and is considerably smaller. The high-density structure noted in Sect.~\ref{sec:3dmods} is visible as a high-radiance structure, appearing to be ring-like from the face-on perspective and irregular in the edge-on view. This asymmetry can also be seen in the tail-on projection, where the high-radiance structure is again tilted by $30\si{\degree}$ to the vertical axis. The irregular structure of the high-density object is, however, not readily apparent \citep[][Sect.~4.4.4]{baalmann_phd}. 

\section{Summary}\label{sec:summ}

Four models of \textlambda-Cephei-like astrospheres with different inflow speeds of the ISM, $u_{\mathrm{ISM}}\in\{80, 40, 20, 10\}$, were computed in both 2D and 3D. The 2D models were simulated without a SW magnetic field and an ISM magnetic field parallel to the ISM inflow, whereas the 3D models were simulated with a simplified Parker spiral for the SW and an oblique field for the ISM. The astrospheric evolution was analysed based on the example of the 2D model with $u_{\mathrm{ISM}}=20\,\si{km/s}$. Synthetic observations in H\textalpha\ were generated for all models. 

It has been found that, due to the dominance of hydrodynamic effects over the magnetic field, the 2D and 3D models are in good agreement for $u_{\mathrm{ISM}}\in\{80, 40\}\,\si{km/s}$, and that the 3D model is reasonably symmetric about its central axis; this is also apparent in the projections. However, the 2D models generated a vacillating MD and astrotail, which is inhibited by the greater influence of numeric diffusion in the more coarsely refined 3D models.

For the models with $u_{\mathrm{ISM}}=20\,\si{km/s}$, where the ISM inflow is barely superfast and supersonic, this symmetry is broken; the 3D model astrosphere is distorted by the oblique ISM magnetic field. However, with the exception of this distortion, the 2D and 3D models still agree remarkably well. At this ISM inflow speed the evolution of the model contains a sizable double-BS structure for a significant time, which is apparent in the synthetic observations. The physical time until the simulation has reached stationarity typically exceeds the lifetime of the star. 

The 2D and 3D models with $u_{\mathrm{ISM}}=10\,\si{km/s}$, where the ISM inflow is subfast and subsonic and therefore does not generate a BS, are not in good agreement. The extent of the astrotail is much smaller in the 3D model, and a high-density structure develops at its downwind end. This structure is highly asymmetric and irregular; it dominates the synthetic H\textalpha\ observations. The origin of this structure lies in the oblique magnetic field; it cannot be ruled out that such a structure is common at the astrotails of model astrospheres either with or without a BS. 

The following conclusions can be summarised:
\begin{enumerate}
    \item 2D MHD simulations can be sufficient to produce accurate models of the large-scale astrospheric structure if hydrodynamic effects are dominant over the ISM magnetic field (Sect.~\ref{sec:3dmods}). However, no SW magnetic field can be simulated with this approach.
    \item Fine-resolution models, which are less computationally expensive in 2D compared to 3D, feature a vacillating astrotail (Sect.~\ref{sec:2dmods}). This is inhibited in coarse-resolution models due to the increased impact of numeric diffusion (Sect.~\ref{sec:3dmods}).
    \item Astrospheres around high-mass stars with comparably slow ISM inflows require extensive simulation times to reach stationarity, potentially much longer than the actual lifetimes of the modelled stars (Sect.~\ref{sec:evol}). This implies that observed astrospheres of such short-lived stars must be considered to be in a transient state.
    \item The model with an ISM inflow too slow to generate a BS shows significant differences between its 2D and 3D simulations (Sect.~\ref{sec:3dmods}). 
    \item Within this model, the oblique magnetic field has generated a hitherto unobserved high-density structure at the downwind end of the astrotail, which dominates the synthetic H\textalpha\ radiance map (Sect.~\ref{sec:synthobs}).
\end{enumerate}

To conclude, the use of 2D MHD can be expedient with regard to its more computationally expensive 3D counterpart as long as the importance of the stellar magnetic field and of the orientation of the ISM magnetic field is negligible. While the 2D approach predetermines cylindrical symmetry, additional features that stem from the finer resolution, such as a vacillating astrotail, can emerge at no additional computational cost with regard to the 3D approach. When modelling in 3D, it can prove vital to extend the simulation boundary to include the full astrotail in order to reproduce all important features of the model.

\begin{acknowledgements}
This work makes use of the ColorBrewer colour scales, designed by Cynthia A.~Brewer, Geography,
Pennsylvania State University (\url{www.ColorBrewer.org}). KS is grateful to the \textit{Deutsche Forschungsgemeinschaft} (DFG), funding the project SCHE334/9-2. JK acknowledges financial support through the \textit{Ruhr Astroparticle and Plasma Physics (RAPP) Center}, funded as MERCUR project St-2014-040. 
\end{acknowledgements}

\bibliographystyle{aa}
\bibliography{2021paper}

\end{document}